# Geometric π Josephson junction: Current-phase relations and critical current

**Andreas Gumann, Christian Iniotakis, and Nils Schopohl**

***Abstract*—Josephson junctions with an intrinsic phase shift of π, so-called π Josephson junctions, can be realized by a weak link of a *d*-wave superconductor with an appropriate boundary geometry. A model for the pairing potential of an according weak link is introduced which allows for the calculation of the influence of geometric parameters and temperature. From this model, current-phase relations and the critical current of the device are derived. The range of validity of the model is determined by comparison with selfconsistent solutions.**

## I. Introduction

If included in a closed superconducting loop, π Josephson junctions lead to spontaneous currents and a degenerate current ground state [1]. If combined with standard Josephson junctions, promising new possibilities for superconducting electronics arise. Complementary superconducting quantum interference devices (SQUIDs) follow from the combination of a standard and a π Josephson junction [2] and can for example be used in order to improve rapid single flux quantum (RSFQ) logic [3] in various ways [4], [5].

Recently, a novel realization of a π Josephson junction based on the boundary geometry of a weak link of an unconventional *d*-wave superconductor has been proposed [6]. This geometric π Josephson junction consists of a strip of an epitactic *c*-axis oriented thin film of a *d*-wave superconductor, which is narrowed down from one side by a wedge-shaped incision (cf. Fig. 1). It has been shown that a transition to negative critical currents indicating π Josephson junction behavior occurs if (1) the crystal orientation is appropriately chosen and (2) the residual width *w* of the weak link is sufficiently small. The *d*-wave superconductor can be a cuprate high-temperature superconductor [7] as well as any other superconductor with *d*-wave symmetry [8]. Because of the simple planar geometry based on a single superconducting layer, the device is highly suitable for circuitry consisting of many junctions and for close packaging on a single substrate. In particular, SQUID and superconducting quantum interference filter (SQIF, [9]) geometries consisting of several geometric π Josephson junctions can be designed in a straightforward way [6].

In the present work, we introduce a model for the pairing potential of a geometric π Josephson junction. This model allows for the characterization of the device based on current-phase relations and critical currents. Applying microscopic Eilenberger equations of superconductivity, the influence of geometric parameters and temperature can be obtained from the model introduced here. In order to determine the range of validity of this model, we compare the results to selfconsistent solutions and quantitatively calculate the critical width *w* for which a transition to the π state occurs.

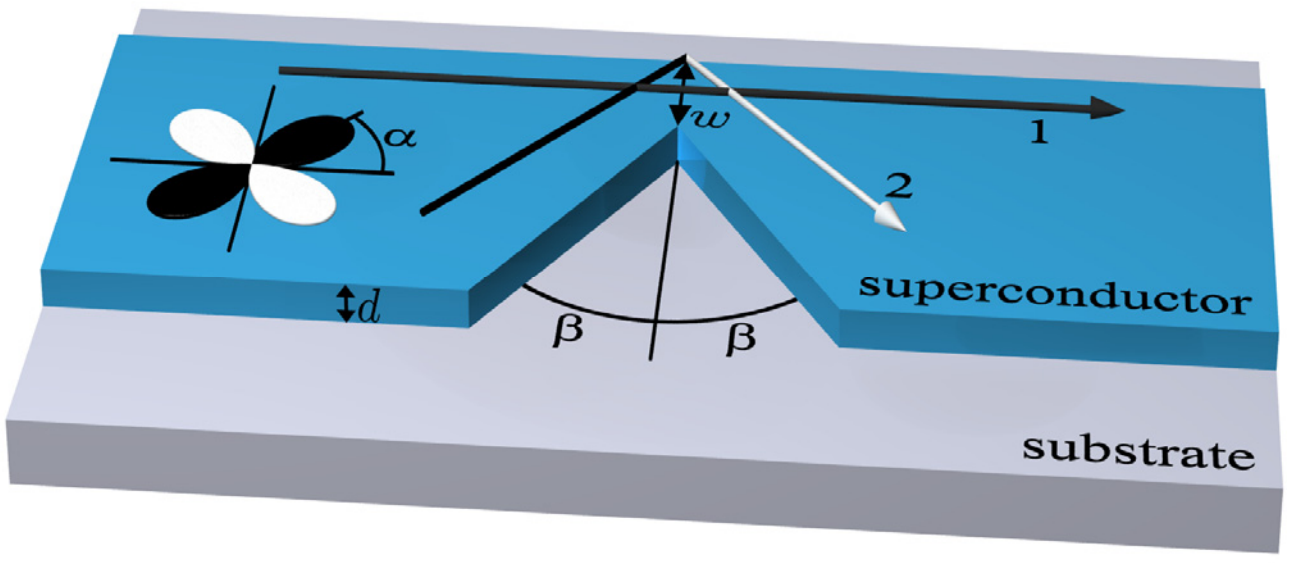

Fig. 1. (Color online) Boundary geometry defining the geometric π Josephson junction based on an epitactic thin film of a *d*-wave superconductor. Two typical quasiparticle trajectories traveling across the junction have been sketched: 1-without reflection, 2-with reflection.

## II. Functional Principle Of The Device

The intrinsic phase shift of the device is a direct consequence of the *d*-wave pairing symmetry. If the residual width of the junction *w* is large, trajectories without and with a reflection at the straight boundary contribute to the total current across the junction (trajectories of type 1 and 2 in Fig. 1). If the constriction is narrow enough, the dominant contribution to the total current through the constriction stems from quasiparticles which get reflected at the straight boundary line opposite to the wedge (type 2). If the orientation of the *d*-wave $\alpha$ is chosen to be $\alpha=\pi/4$, the reflected quasiparticles suffer a sign change of the pairing potential which leads to the formation of Andreev bound states and a phase shift.

## III. Theoretical Description

In order to calculate current-phase relations and critical currents, we solve microscopic Eilenberger equations of superconductivity [10], [11]. We assume a cylindrical Fermi

A. Gumann and N. Schopohl are with Institute for Theoretical Physics and Center for Collective Quantum Phenomena, University of Tuebingen, Auf der Morgenstelle 14, D-72076 Tübingen, Germany (phone: +49-(0)-7071-29-76098; fax: +49-(0)-7071-29-5604; e-mail: andreas.gumann@uni-tuebingen.de, nils.schopohl@uni-tuebingen.de ).

C. Iniotakis is with Institute for Theoretical Physics, ETH Zurich, Schafmattstrasse 32, CH-8093 Zurich, Switzerland (email: iniotakis@itp.phys.ethz.ch ).

surface with the cylinder axis aligned perpendicular to the film plane. Accordingly, the Fermi velocity is given by $\mathbf{v}_F = v_F(\hat{\mathbf{x}}\cos\theta + \hat{\mathbf{y}}\sin\theta)$ with $\hat{\mathbf{x}},\hat{\mathbf{y}}$ being unit vectors in the film plane and $\theta$ being the polar angle. Then, the self-consistency equation for the position-dependent part $\Delta(\mathbf{r})$ of the *d*-wave pairing potential $\Delta(\mathbf{r},\mathbf{k}_F) = \Delta(\mathbf{r})\cos(2\theta - 2\alpha)$ reads:

$$\Delta(\mathbf{r}) = 2\pi N(0)Vk_BT\sum_{\varepsilon_n>0}^{\omega_c}\left\langle f(\mathbf{r},\mathbf{k}_F,i\varepsilon_n)\cos(2\theta-2\alpha)\right\rangle_\theta \ . \tag{1}$$

Here, $N(0)$ is the normal density of states at the Fermi surface, $V$ is the coupling constant, $\varepsilon_n = (2n+1)\pi k_B T$ are Matsubara frequencies , and $\langle\ldots\rangle_\theta$ denotes Fermi surface averaging.

Once, the pairing potential $\Delta(\mathbf{r})$ has been calculated selfconsistently, the current density follows from

$$\mathbf{j}(\mathbf{r}) = 4\pi eN(0)k_BT\sum_{\varepsilon_n>0}^{\omega_c}\left\langle \mathbf{v}_F\ g(\mathbf{r},\mathbf{k}_F,i\varepsilon_n)\right\rangle_\theta \ . \tag{2}$$

Details of the methods employed in order to calculate selfconsistent solutions can be found in [6], [12].

For comparison, we introduce a non-selfconsistent model for the pairing potential $\Delta(\mathbf{r},\mathbf{p}_F)$. This model assumes a step-like variation of the phase of the pairing potential, whereas its amplitude is taken to be constant:

$$\Delta_{L,R}(\mathbf{r},\mathbf{p}_F) = \Delta_\infty(T)\cos(2\theta-2\alpha)\,e^{\mp i\gamma/2} \ . \tag{3}$$

Here, the indices $L,R$ label the left and right side of the junction, $\Delta_\infty(T)$ is the temperature-dependent amplitude of the pairing potential in the bulk, and $\gamma = \phi_R - \phi_L$ is the phase difference across the junction. Using this model, the current flowing through the junction can be calculated analytically from the current equation (2).

## IV. Results

### A. Current-Phase Relations

In Fig. 2, we show current-phase relations obtained from the step model (1) for different widths $w$ of the junction as well as for different temperatures. The width $w$ is given in terms of the coherence length $\xi_0 = \hbar v_F/\pi\Delta_\infty(T=0)$. For the step model, the opening angle of the wedge is $\beta=0$ and we use the orientation angle $\alpha=\pi/4$ which corresponds to the most pronounced occurrence of the π state.

For very small values of the width of the junction, the currents are negative for all values of the phase difference γ at all temperatures $T$. Close to $T_c$, the current-phase relations assume the form $I = I_c\sin(\gamma+\pi) = -I_c\sin\gamma$. Close to $T=0$, however, the current-phase relations approach a saw-tooth-like behavior. In the limit $w\to 0$, the special case of a π point contact is being realized, in direct analogy to a standard point contact but with an intrinsic phase shift of π.

With growing width of the junction, the currents become positive. The transition to positive currents first occurs at temperatures close to $T_c$. At lower temperatures, the currents remain negative, predominantly for small values of the phase difference γ. At high temperatures, the current-phase relations always assume a highly sinusoidal form, whereas at low temp-eratures, higher harmonics play a dominant role. For large values of the width $w$, the currents are positive for all phase differences $\gamma$ at all temperatures $T$.

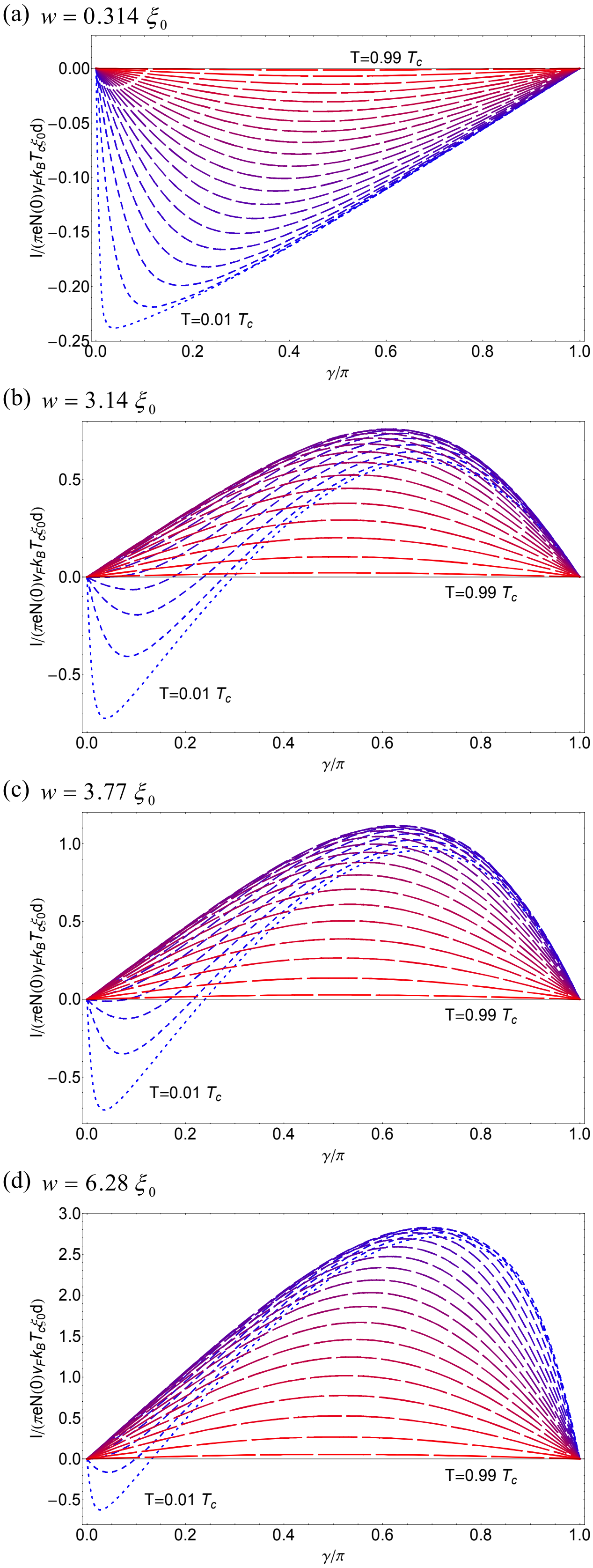

Fig. 2. (Color online) Current-phase relations obtained from the step model (1). In the four subfigures, four different values of the width $w$ of the junction have been used (as indicated). Within each subfigure, current-phase relations for temperatures $T$ = 0.01, 0.05, 0.1, …, 0.9, 0.95, 0.99 $T_c$ (corresponding to growing length of the dashes) are shown

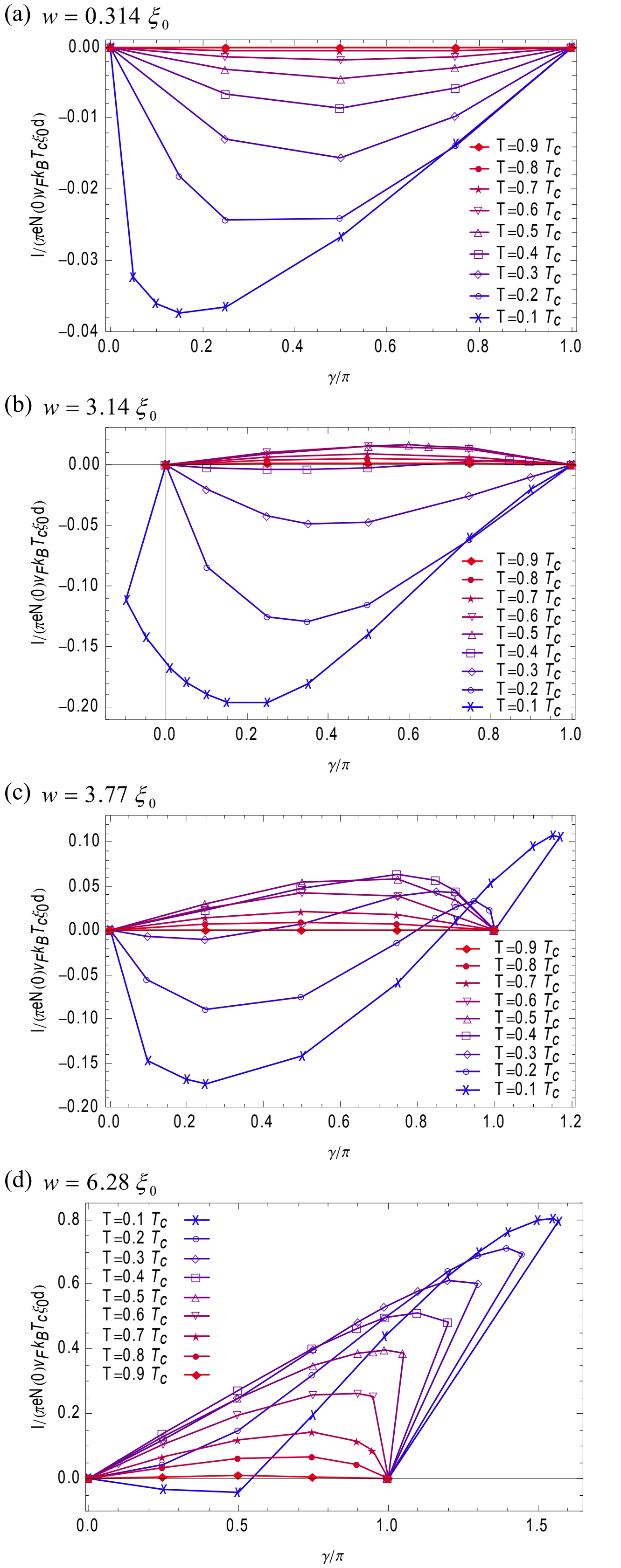

Fig. 3. (Color online) Selfconsistently determined current-phase relations. In the four subfigures, four different values of the width $w$ of the junction have been used. Within each subfigure, current-phase relations for temperatures $T$ = 0.1, 0.2, …, 0.8, 0.9 $T_c$ (as indicated) are shown.

In Fig. 3, we show selfconsistently calculated current-phase relations for the same four different values of the width $w$ as in Fig. 2. As in Fig. 2, we use $\beta=0$ and $\alpha=\pi/4$.

In the case of the selfconsistent calculations, the absolute values of the currents are much smaller. This is due to the fact that local suppression of the amplitude of the pairing potential is considered in the selfconsistent calculations. In contrast, the amplitude of the pairing potential is assumed to be constant in the step model. The general form of the current-phase relations is similar to the results of the step model, but a very distinct difference exists. In the case of the step model, the current-phase relations are always single-valued, whereas, in the case of the selfconsistent calculations, multi-valued current-phase relations are possible.

## B. *Critical current*

The critical current of a Josephson junction is given by the absolute maximum of the current-phase relation. In Fig. 4, we show the critical current $I_c(w,T)$ for the step model. In subfigure (a), the data are plotted for fixed width $w$ as a function of $T$, whereas, in (b), the same data are plotted for fixed $T$ as a function of $w$. The two plots correspond to two experimental situations which can be thought of to verify the 0-π transition, i.e. by variations of $T$ and $w$, respectively. The results in (a) show that the π state (indicated by a negative critical current) is predominantly being entered at low temperatures. For very small values of $w$, however, the π state survives up to $T_c$. From (b), we find that at $T$=0.1 $T_c$, the critical width $w$ is about $\approx 2.5\xi_0$. With increasing temperature, the critical width $w$ of the 0-π transition is being shifted to smaller values of $w$.

In Fig. 5, we show corresponding results for the critical current $I_c(w,T)$ from the selfconsistent calculations. The general behavior found from the step model can also be found in the selfconsistent results. However, some peculiar differences occur.

First, the absolute values of the critical current are overestimated in the step model. Second, close to $T_c$, the critical current for a fixed width $w$ as a function of temperature exhibits a linear increase with decreasing temperature in the step model. From the selfconsistent solutions, however, we find deviations from this linear behavior. Third, the critical width found from the step model is smaller than from the selfconsistent result. In the selfconsistent solutions, a critical width of about $w\approx 4\xi_0$ follows at $T$=0.1 $T_c$. All these three differences between the step model and the selfconsistent results are due to the neglect of local suppression of the pairing potential in the step model.

Further calculations based on the step model (not shown here) indicate that the orientation of the crystal lattice $\alpha$ does not have critical influence on the operation of the Josephson device. Small deviations from the ideal alignment $\alpha=\pi/4$ do not lead to a disappearance of the π state. Using the self-consistent calculations, it has been found that the opening angle of the wedge $\beta$ does not have critical influence, either.

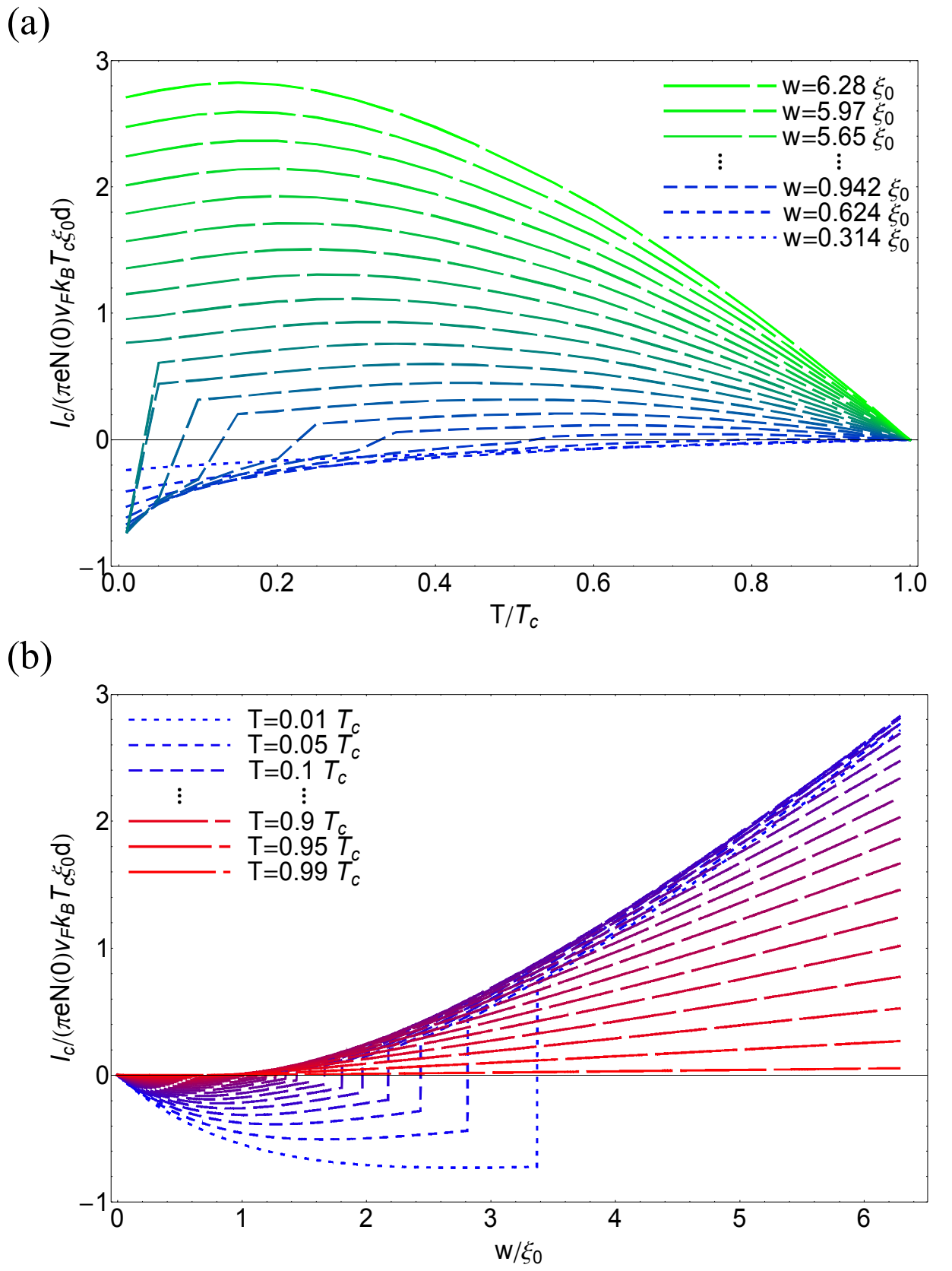

Fig. 4. (Color online) Critical current $I_c(w,T)$ from the step model (1) for $\beta=0$ and $\alpha=\pi/4$.

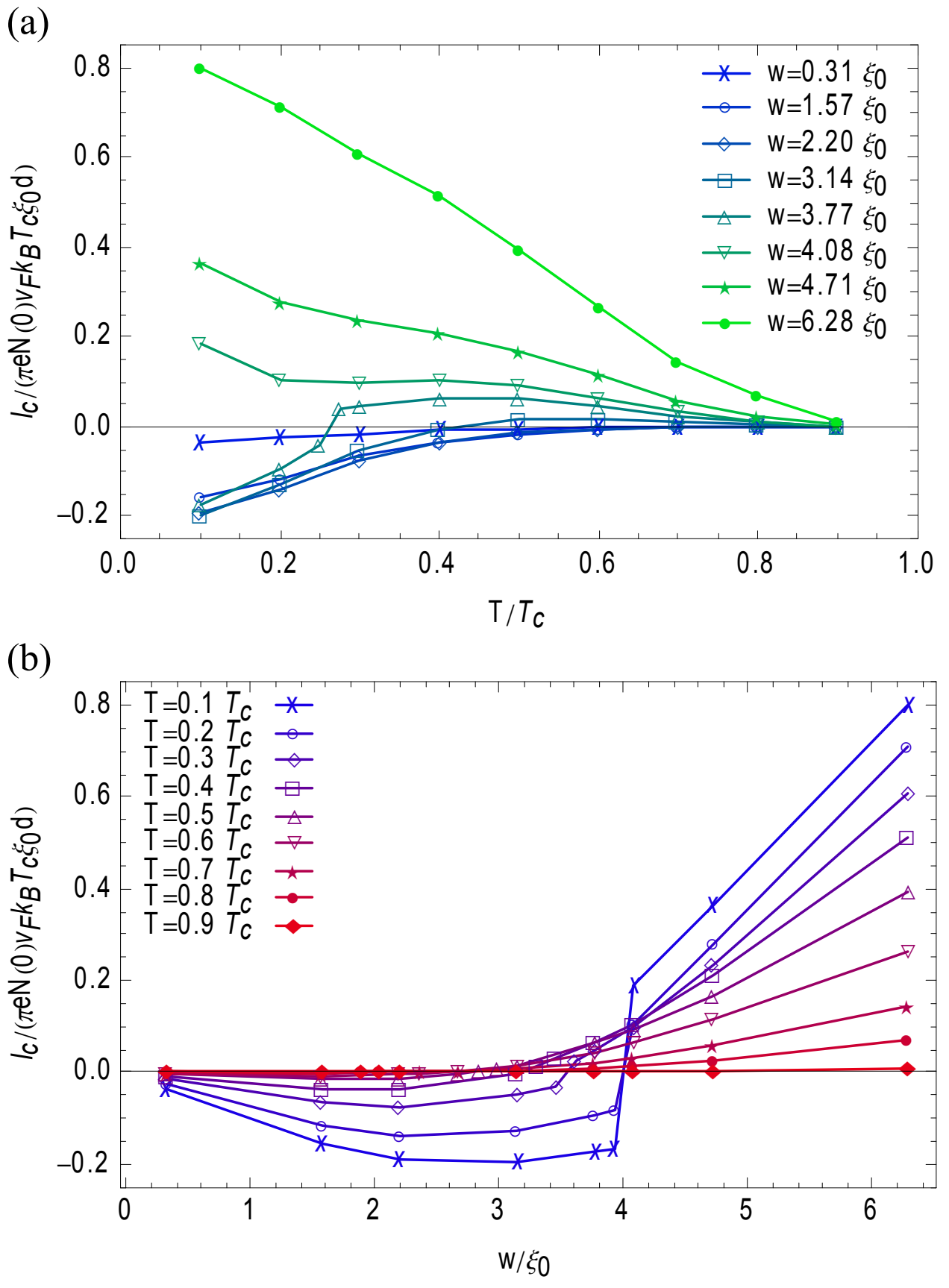

Fig. 5. (Color online) Selfconsistent critical current $I_c(w,T)$ for $\beta=0$ and $\alpha=\pi/4$.

For growing values of $\beta$, reflected quasiparticles from a reduced angular interval contribute to the phase shift, but for values up to $\beta=\pi/4$, the critical width $w$ remains almost unaltered.

## V. Discussion And Conclusions

The step model for the pairing potential of a geometric π Josephson junction introduced in the present work has been used in order to calculate current-phase relations and the critical current of the device. By comparison with selfconsistent solutions, we find that the step model provides qualitatively correct behavior whereas only selfconsistent solutions can be used for quantitative predictions.

The current-phase relations of the geometric π Josephson junction considered in the present work in general strongly deviate from the standard sinusoidal form. Predominantly at low temperatures, higher harmonics play a dominating role. Therefore, we expect interesting behavior also in the resistive mode. The current-phase relations are directly connected with the Shapiro steps occurring under microwave irradiation. Accordingly, we expect that the Shapiro steps should deviate from the standard form and possibly allow for a confirmation of the transition to the π state.